\documentclass[aps,pra,twocolumn,showpacs,groupedaddress]{revtex4}
\usepackage{graphicx} 
\usepackage{amsmath}
\usepackage{longtable}

\begin{document}

\title{Positronic complexes with unnatural parity}
\author{M.W.J.Bromley}
  \email{mbromley@physics.sdsu.edu}
\affiliation{Department of Physics and Computational Sciences Research Center, San Diego State University, San Diego CA 92182, USA}
\author{J.Mitroy}
  \email{jxm107@rsphysse.anu.edu.au}
\affiliation{Faculty of Technology, Charles Darwin University, Darwin NT 0909, Australia}
\author{K.Varga}
  \email{kalman.varga@Vanderbilt.Edu}
\affiliation{Department of Physics and Astronomy, Vanderbilt University, Nashville, Tennessee 37235, USA}

\date{\today}

\begin{abstract}

The structure of the unnatural parity states of PsH, LiPs, NaPs and 
KPs are investigated with the configuration interaction and stochastic
variational methods.  The binding energies (in hartree) are found to be 
$8.17 \times 10^{-4}$, $4.42 \times 10^{-4}$, $15.14 \times 10^{-4}$ and 
$21.80 \times 10^{-4}$ respectively.  These states are constructed
by first coupling the two electrons into a configuration which is 
predominantly $^3$P$^{\rm e}$, and then adding a $p$-wave positron.  
All the active particles are in states in which the relative angular
momentum between any pair of particles is at least $L = 1$.  The
LiPs state is Borromean since there are no 3-body bound subsystems 
(of the correct symmetry) of the (Li$^+$, $e^-$, $e^-$, $e^+$) 
particles that make up the system.  The dominant decay mode of these  
states will be radiative decay into a configuration that autoionizes 
or undergoes positron annihilation.  

\end{abstract}

\pacs{36.10.-k, 36.10.Dr, 34.85.+x}

\maketitle 

\section{Introduction} 

The stability of a bound state composed of two electrons and a
positron, the positronium negative ion, was first demonstrated
in a seminal calculation by Wheeler \cite{wheeler46}.  Shortly 
after this calculation, the four body systems, PsH and Ps$_2$ 
were shown to be stable \cite{hylleraas47b,ore51}.  Since that
time, only a few other electronically stable states have been
discovered that can be formed from combinations of $p^+$,$e^-$ 
and $e^+$.  These are additional bound states of Ps$_2$ 
\cite{kinghorn93,varga98a,usukura00a,schrader04a}, a compound that is best 
described as $e^+$PsH \cite{varga99a}, and a ($p^+$, 4$e^-$, 2$e^+$) 
complex \cite{varga99a}.  Additionally, a number of atoms 
have been identified as being capable of binding positronium
and positrons \cite{ryzhikh97a,strasburger98,mitroy02b}      

Just recently, a new class of positronic compounds that are 
electronically stable was identified \cite{mitroy07b}.  
The new PsH and NaPs bound states were unnatural parity 
states  with symmetry conditions that act to prevent decay 
into the lowest energy dissociation products.  An unnatural 
parity state is a state with parity equal to 
$\Pi = (-1)^{^{L+1}}$ where $L$ is the orbital angular 
momentum of the state.   These PsH and NaPs systems 
have the two valence electrons in a spin-triplet state, a 
total orbital angular momentum of zero, and an odd parity, 
i.e. $L^{\Pi} = 0^-$. In addition, these states had the 
unusual feature of decaying very slowly by $2\gamma$ or 
$3\gamma$ annihilation.  

In this paper, more details about the $L^{\Pi} = 0^-$
negative parity states of PsH and NaPs are given.  Negative
parity states of LiPs and KPs are also identified as being
electronically stable.  The LiPs state has the additional
distinction of being a Borromean state
\cite{zhukov93a,blume02a,richard03a,richard06a} 
since the (Li$^+$, $e^-$, $e^-$, $e^+$) system has no stable 
3-body state that can act as a parent for the four body 
$^{2,4}$S$^{\rm o}$ state.  We have adopted the definition of 
Richard: {\em A bound state is Borromean 
if there is no path to build it via a series of stable states by 
adding the constituents one by one} \cite{richard03a}.   

It should be noted that there are analogs of these states in
the alkaline-earth sequence.  Configuration interaction (CI) 
approaches has been used to demonstrate the stability of the 
Be$^-$, Mg$^-$, Ca$^-$ and Sr$^-$ $np^3$ $^4$S$^{\rm o}$ states 
\cite{bunge82a,hart99a}.  However, the issue of whether an electron 
can be attached to the $^3$P$^{\rm e}$ state of H$^-$ into an 
$^4$S$^{\rm o}$ state of H$^{2-}$ has been the subject of some 
controversy.  A complex rotation method was applied to a large 
basis CI wave function and a shape threshold lying about 1.4 eV 
above the $^3$P$^{\rm e}$ threshold was predicted \cite{sommerfeld97a}.  
However, this was contradicted by a much more sophisticated 
hyperspherical calculation that exhibited no sign of a resonance 
\cite{morishita98a}. 

\section{Theoretical Overview} 

\subsection{Symmetry conditions for binding}  

The stability of these systems lies in the symmetry relations
between the pairs of particles that make up the system.  
The discussion of these conditions will be addressed 
specifically to PsH, but these conditions, with some small
modifications, will also apply to the other systems addressed
in this paper.

The electronic stability of PsH can be motivated by consideration of the 
H$^-$($2p^2$ $^3$P$^{\rm e}$) bound state \cite{holoien61a,drake70a,bylicki03a}.  
This state has an energy of $-0.12535545$ hartree \cite{bylicki03a} and is 
electronically stable due to symmetry conditions.  It cannot decay
into the H($1s$) + $e^-$ channel since the $\ell = 1$ partial wave of
the electron automatically results in a state of negative parity.  
The $L^{\Pi} = 0^-$ state of PsH is formed when the positron is 
trapped into a $2p$ state of the H$^-$ attractive potential well.  
The possible decay modes are constrained by the symmetry conditions.
Dissociation into Ps($1s$)+H($1s$) is forbidden since $\Pi = (-1)^L$ 
where $L$ is the orbital angular momentum between the Ps($1s$) and 
H($1s$) fragments.  Similarly, dissociation into Ps($ns$)+H($n\ell$) 
or Ps($n\ell$)+H($ns$) does not occur since it is not possible 
to construct an $L^{\Pi} = 0^{-}$ state if one of the angular 
momentum is zero.  The lowest energy dissociation channel would 
be into Ps($2p$)+H($2p$) (with the two fragments in a $p$-wave) 
with an energy threshold of $-0.1875$ hartree.  Another possible 
decay would be into the H$^-$($2p^2$ $^3$P$^e$) + $e^+$ channel 
but the threshold energy here is  $-0.12535545$ hartree 
\cite{bylicki03a}.   

It is easy to see that there is potentially a large energy advantage 
associated with binding the positron to the negative ion.   If the 
H$^-$ state is regarded as a point particle with an internal energy 
of $\approx -0.125$ hartree, then a positron in the $2p$ state will 
lower the total energy to $-0.250$ hartree.  In actuality the 
H$^-$($2p^2$ $^3$P$^{\rm e}$) state is diffuse \cite{drake70a}, but 
the advantage of attaching the positron to the negative ion is clear.

\subsection{Symmetry conditions for annihilation}  

The dominant electron-positron annihilation processes are 
the $2\gamma$ and $3\gamma$ processes.  The 2$\gamma$ 
annihilation rate for bound systems is proportional to the 
probability of finding an electron and a positron at the
same position in a spin-singlet state according to
\begin{eqnarray}
\Gamma &=&  4\pi r_e^2 c \langle \Psi |
      \sum_{i} O^S_{ip} \delta( \mathbf{r}_i - \mathbf{r}_p ) | \Psi \rangle \nonumber \\
       &=& 2.018788 \times 10^{11} \sum_{i}
        \langle \delta( \mathbf{r}_i - \mathbf{r}_p ) \rangle_S \ ,
\label{Gamma1}
\end{eqnarray}
\cite{lee58a,ryzhikh99a,neamtan62},
where the sum is over the electron coordinates, the $\delta$-function
expectation is evaluated in $a_0^3$, and $\Gamma$ is given numerically
in s$^{-1}$.  The operator $O^S_{ip}$ is a spin projection operator
to select spin-singlet states for the $i,p$ electron-positron pairs
in the wave-function $\Psi$ which is anti-symmetrized in the
electron coordinates.  The rate constant for the Ps ground state 
is about $8 \times 10^9$ s$^{-1}$.
Equation (\ref{Gamma1}) involves a contact interaction which means 
that the relative angular momentum of the annihilating pair 
($L_{\rm rel}$) must be zero \cite{lee58a}.  However, 
electron-positron annihilation is possible even for when
the relative angular momentum of the annihilating pair is greater 
than zero.  For example the Ps($2p$) levels can undergo $2\gamma$ 
annihilation at rates proportional to $\alpha^5$ and $\alpha^6$ 
respectively \cite{alekseev58a,alekseev59b}.  
The rates for the different Ps($2p$) levels have been calculated 
to be approximately 10$^4$ s$^{-1}$ \cite{alekseev58a,alekseev59b}.
Similarly, the $3\gamma$ process, which happens when the
annihilating pair are in a spin-triplet state, can also
occur at a rate proportional to $\alpha^6$ when the pair have 
a relative angular momentum of 1.  The discussions below
about the symmetry conditions for positron annihilation
concern the fast $2\gamma$ and $3\gamma$ processes for
pairs in relative $s$-states.

Consider the electron-positron annihilation of a PsH state 
of $^{2}$S$^{\rm o}$ symmetry.  The relative angular momentum of
the annihilating pair ($L_{\rm rel}$) must be zero.   This means 
the total angular momentum of the state will come from 
the center-of-mass motion of the annihilating pair ($L_{\rm cm}$), 
and from the angular momentum of the spectator electron 
($L_{\rm spectator}$).   The total parity of the state is 
determined by the parity of the individual constituents, i.e.
$\Pi = (-1)^{L_{\rm spectator}+L_{\rm cm}+L_{\rm rel}}$.  It 
is simply not possible to form an odd parity state with a total 
angular momentum of zero if any one of the angular 
momenta is zero.  Consequently, a two electron/one positron 
state of $^{2}$S$^{\rm o}$ symmetry cannot decay by 
the fast $2\gamma$ process.

These arguments also apply to the $3\gamma$ annihilation
process.   The $3\gamma$ process occurs for electron-positron
pairs in a spin-triplet state with a relative angular momentum
of zero.  Once again, it is simply impossible to form a state
of $^{2}$S$^{\rm o}$ (or $^{4}$S$^{\rm o}$) symmetry if the
relative angular momentum of the annihilating pair is zero.
So it is reasonable to conclude that the lowest order $3\gamma$ 
decay is not possible from a $^{2,4}$S$^{\rm o}$ state.  

\section{Calculation methods} 

\subsection{The configuration interaction method} 

A majority of the calculations in the present paper were performed 
with a configuration interaction approach 
\cite{bromley02a,bromley02b,mitroy06a}.  The CI basis was 
constructed by letting the two electrons (particles 1 and 2) 
and the positron (particle 0) form all the possible total 
angular momentum $L_T = 0$ configurations, with the two 
electrons in a spin-triplet state, subject to the selection rules,
\begin{eqnarray}
\max(\ell_0,\ell_1,\ell_2) & \le & J \ , \\
\min(\ell_1,\ell_2)& \le & L_{\rm int} \ ,  \\  
(-1)^{(\ell_0+\ell_1+\ell_2)}& = & -1  \ . 
\end{eqnarray}
In these rules $\ell_0$, $\ell_1$ and $\ell_2$ are respectively 
the orbital angular momenta of the positron and the two electrons.  
We define $\langle E \rangle_J$ to be the energy of the calculation 
with a maximum orbital angular momentum of $J$.  The single
particle orbitals were Laguerre Type Orbitals (LTOs) with a 
common exponent chosen for all the orbitals of a common $\ell$  
\cite{bromley02a,bromley02b,mitroy06a}.  The orbital basis sets 
for the positron and electrons were identical.      

A major technical problem afflicting CI calculations 
of positron-atom interactions is the slow convergence of the energy 
with $J$ \cite{mitroy02b,mitroy06a}.  The  $J \rightarrow \infty$ 
energy, $\langle E \rangle_{\infty}$, is determined by the use of 
an asymptotic analysis.  The successive increments, 
$\Delta E_{J} = \langle E \rangle_J - \langle E \rangle_{J-1}$, 
to the energy can be written as an inverse power series 
\cite{schwartz62a,carroll79a,hill85a,mitroy06a,bromley07a}, viz 
\begin{equation}
\Delta E_J \approx \frac {A_E}{(J+{\scriptstyle \frac{1}{2}})^6} 
    + \frac {B_E}{(J+{\scriptstyle \frac{1}{2}})^7} 
    + \frac {C_E}{(J+{\scriptstyle \frac{1}{2}})^8} 
    + \frac {D_E}{(J+{\scriptstyle \frac{1}{2}})^9} + \dots \ \   .
\label{extrap1}
\end{equation}
The first term in the series starts with a power of 6 since
all the possible couplings of any two of the particles result
in unnatural parity states \cite{kutzelnigg92a}.    

The $J \to \infty$ limit, has been determined by fitting sets of 
$\langle E \rangle_J$ values to asymptotic series with either 1, 2, 3  
or 4 terms.  The coefficients, $A_E$, $B_E$, $C_E$  and $D_E$ for the 4-term 
expansion are determined at a particular $J$ from 5 successive energies
($\langle E \rangle_{J-4}$,  $\langle E \rangle_{J-3}$, 
$\langle E\rangle_{J-2}$, 
$\langle E \rangle_{J-1}$ and $\langle E \rangle_{J}$).  Once the 
coefficients have been determined it is easy to sum the series to 
$\infty$ and obtain the variational limit.   Application of asymptotic 
series analysis to helium has resulted in CI calculations reproducing 
the ground state energy to an accuracy of $\approx \!\! 10^{-8}$ hartree
\cite{salomonson89b,bromley07a}.  

The treatment of the alkalis Li, Na and K  
requires the use of a frozen core approximation.  The
details of this approximation have been discussed in great
detail elsewhere \cite{bromley02a,bromley02b,mitroy06a}, 
so only the briefest description is given here.  The model 
Hamiltonian is initially based on a hartree-Fock (HF) wave 
function for the neutral atom ground state.  The core orbitals
are then frozen.  The direct part of the core potential is 
attractive for electrons and repulsive for the positron.
The impact of the direct and exchange part of the HF core
interactions on the active particles are computed without 
approximation.  One- and two-body semi-empirical polarization 
potentials are then added to the potential.  The adjustable 
parameters of the core-polarization potential are defined by 
reference to the spectrum of neutral atom 
\cite{bromley02b,mitroy03f}.

\subsection{The stochastic variational method (SVM)} 

In the stochastic variational approach an explicitly correlated
gaussian (ECG) is constructed by placing the particles
(electrons and positrons) into Gaussian single particle orbitals
\begin{equation}
r_i^l Y_{lm}({\hat r_i}) {\rm exp}\lbrace-\beta_i r_i^2\rbrace=
{\cal Y}_{lm}({\bf r}_i) {\rm exp}\lbrace-\beta_i r_i^2\rbrace \ , 
\end{equation}
and using a
\begin{equation}
{\rm exp}\lbrace-\alpha_{ij} ({\bf r}_i-{\bf r}_j)^2\rbrace \ , 
\end{equation}
Gaussian correlation function between the $i$th and $j$th particles. 
The $N$-particle trial function is then
\begin{eqnarray}
\Psi_{LS}({\bf r}) & = & \! {\cal A}
\lbrace
\left[
\left[ \left[ {\cal Y}_{l_1} {\cal Y}_{l_2}\right]_{l_{12}}
{\cal Y}_{l_3}\right]{\ldots} \right]_{LM_L} \chi_{SM_S} \nonumber \\ 
& \times & \! \! \prod_{i=1}^N {\rm exp}\lbrace -\beta_i r_i^2\rbrace
\prod_{i<j}   {\rm exp}\lbrace
-\alpha_{ij} ({\bf r}_j-{\bf r}_i)^2\rbrace \rbrace
\label{SVML}
\end{eqnarray}
where ${\cal A}$ is an antisymmetrizer and $\chi_{SM_S}$ is the spin
function of the particles.  The nonlinear variational parameters 
$\beta_i$ and $\alpha_{ij}$ are selected by an iterative trial and
error procedure.  Full details are given in Ref.~\cite{suzuki98a,suzuki98b}.
The orbital angular momentum quantum numbers $l_i$ are restricted to 
occupy the lowest possible values consistent with the overall 
symmetry of the state.  The spherical part of the ECG basis functions 
effectively allows internal angular momentum to be distributed between 
the different parts of the systems.  Accordingly, eq.~(\ref{SVML}) 
implicitly includes all possible internal symmetries that can make a 
contribution to the energy.  This has been verified with test 
calculations.
 
\section{Results of calculations} 

\subsection{The A$^-$($np^2$ $^3$P$^{\rm e}$) states} 

\begin{table}[th]
\caption[]{  \label{tab0}
The energies of various parent states relevant to the structure 
and energy threshold for the APs unnatural parity states. The 
polarizability only allows for $np \to kd$ excitations since
$np \to ks$ excitations cannot occur in the $^3$P$^{\rm e}$ 
channel.  The energy for the H$^-$($np^2$ $^3$P$^{\rm e}$) state
was taken from Bylicki and Bednarz \cite{bylicki03a}, while
those for the alkali systems were from the present CI calculations.    
} 
\begin{ruledtabular}
\begin{tabular}{lcccc} 
    &  A($np$)          &  A($np^2$ $^3$P$^e$) &  A($np$) + Ps($2p$) &  $\alpha_d \ (a_0^3)$ \\ \hline 
H   &  $-0.12500$      &  $-0.12535545$        &  $-0.1875$      & 173.3  \\  
Li  &  $-0.13023850$   &  Unbound              &  $-0.19273850$  & 142.7   \\  
Na  &  $-0.11156287$   &  $-0.11382478$        &  $-0.17406287$  & 302.0   \\  
K   &  $-0.10018265$   &  $-0.10450418$        &  $-0.16268271$  & 557.6  \\  
\end{tabular}
\end{ruledtabular}
\end{table} 

Table \ref{tab0} gives the energies of the various parent 
states of the APs systems.  These are relevant to the
determination of the energy thresholds.  The energies
for the A$^-$($np^2$ $^3$P$^{\rm e}$) states were taken from
CI calculations which used an exact subset of the basis 
used for the calculations upon the APs system.  The energy 
for the H$^-$($np^2$ $^3$P$^{\rm e}$) state was taken from a 
large CI-Hylleraas calculation \cite{bylicki03a} that was
converged to eleven significant digits.  The results of
an SVM calculation of this state are detailed in Table
\ref{SVM}.  
 
The dipole polarizabilities listed in Table \ref{tab0} show an 
interesting correlation between the polarizability of the A($np$) 
state and the electron affinity in the A$^-$($np^2$ $^3$P$^{\rm e}$) 
channel.  The larger the polarizability, the larger the binding
energy.  The Li$(2p)$ level has the smallest polarizability, 
and is the one atom that is unable to support a negative ion
in the $^3$P$^{\rm e}$ channel.

This behaviour is reminiscent of the electron affinity systematics 
of the alkaline-earths in the $^2$P$^{\rm o}$ channel.  The 
ground states of Be and Mg do not have an electron affinity 
while those of Ca, Sr and Ba have electron affinities that 
become larger as the atom, and its polarizability, become larger
\cite{buckman94}.  The critical polarizability for the 
alkali sequence is somewhere between 142 and 173 $a_0^3$.  The
polarizability of calcium, which just binds an electron
with an electron affinity of $\approx 7\times10^{-4}$ hartree 
\cite{nadeau92,walter92,petrunin96}, is about 160 $a_0^3$  
\cite{mitroy03f,porsev02a}.

\begin{table} [th]
\caption[]{ \label{SVM} Properties of the H$^-$($^3$P$^{\rm e}$), 
PsH($^{2,4}$S$^{\rm o}$) and LiPs($^{2,4}$S$^{\rm o}$) ground 
states.  Data are given assuming an infinite nuclear mass ($n$).  
All quantities are given in atomic units.
The magnitude of the binding energy against dissociation into the
lowest energy fragmentation channels is given by $\varepsilon$ while
$T_+$ and $T_-$ represent the positron and electron kinetic
energy operators.   }
\vspace{0.2cm}
\begin{ruledtabular}
\begin{tabular}{lccc}
Property                            & PsH         &  H$^{-}$ &  LiPs \\ \hline
$N$                                 & 400         &    400   &  1000  \\
$\langle V\rangle/\langle T \rangle$ + 2 &
$5.1 \times 10^{-8}$              &  $4.4 \times 10^{-9}$ &  $5.6 \times 10^{-5}$ \\
$E$                                 & $-0.188317$ &  $-0.12535545$  & $-7.472871$ \\
$\varepsilon $                      & 0.000817   &  0.00035545 & $0.000215$ \\
$\langle T_- \rangle$               & 0.156579554 & 0.125355451 &  \\
$\langle T_+ \rangle$               & 0.031737544 &             & \\
$\langle r_{ne^-} \rangle $       & 8.867       & 11.657619   &  4.3177  \\
$\langle r_{ne^+} \rangle $       & 14.243      &             & 12.991  \\
$\langle r_{e^-e^-}\rangle $        & 14.001      & 19.58289    &  7.6429 \\
$\langle r_{e^+e^-}\rangle $        & 12.722      &             & 12.531   \\
$\langle 1/r_{ne^-} \rangle $   & 0.174397    &  0.160521   &  1.4358 \\
$\langle 1/r_{ne^+} \rangle $   & 0.084716    &             &  0.089443 \\
$\langle 1/r_{e^-e^-}\rangle $      & 0.089023    & 0.0700331   &  0.40130  \\
$\langle 1/r_{e^+e^-}\rangle $      & 0.101789    &             &  0.096298 \\
$\langle r^2_{ne^-} \rangle $   & 121.185     & 271.2046    &  45.959 \\
$\langle r^2_{ne^+} \rangle $   & 247.910     &             &  192.691 \\
$\langle r^2_{e^-e^-}\rangle $      & 245.959     & 556.893     &  92.533 \\
$\langle r^2_{e^+e^-}\rangle $      & 202.204     &             &  182.800 \\
\end{tabular}
\end{ruledtabular}
\end{table}

\subsection{PsH} 

\subsubsection{The configuration interaction method} 

\begin{table*}[th]
\caption[]{  \label{tab:PsH}
The energy of the $^{2,4}$S$^{\rm o}$ state of PsH as a function 
of $J$ and with $L_{\rm int} =3$.  The threshold for binding 
is $-0.1875$ hartree. The column $n$ gives the total number of 
occupied electron orbitals (the number of positron orbitals was 
the same) while $N_{CI}$ gives the total number of configurations.  
The radial expectation values for the electron, $\langle r_{e} \rangle$,
and the positron, $\langle r_{p} \rangle$ are given in $a_0$. 
The results of the $J \to \infty$ extrapolations using
eq.~(\ref{extrap1}) at $J = 10$ are given.
}
\begin{ruledtabular}
\begin{tabular}{lcccccc} 
$J$ &  $n$  &   $N_{CI}$ & $\langle E \rangle_{J}$ & $\varepsilon$ & $\langle r_e \rangle$ &  $\langle r_p \rangle$
                                                               \\ \hline
 1  &  20   &    4200  & $-0.16755818$ & $-0.01994182$ &  7.08076 & 13.10807 \\ 
 2  &  40   &   16400  & $-0.17938458$ & $-0.00811542$ &  6.95155 & 11.88797 \\ 
 3  &  60   &   45000  & $-0.18327391$ & $-0.00422609$ &  7.08870 & 11.66425 \\ 
 4  &  80   &   85000  & $-0.18510516$ & $-0.00239484$ &  7.23884 & 11.70642 \\ 
 5  &  100  &   129200 & $-0.18612684$ & $-0.00137316$ &  7.37821 & 11.82951 \\ 
 6  &  120  &   177200 & $-0.18675237$ & $-0.00074763$ &  7.50464 & 11.97767 \\ 
 7  &  140  &   225200 & $-0.18715897$ & $-0.00034103$ &  7.61882 & 12.13043 \\ 
 8  &  160  &   273200 & $-0.18743569$ & $-0.00006431$ &  7.72208 & 12.27939 \\ 
 9  &  180  &   321200 & $-0.18763074$ &  0.00013074 &  7.81562 & 12.42101 \\ 
10  &  200  &   369200 & $-0.18777213$ &  0.00027213 &  7.90047 & 12.55387 \\ \hline  
\multicolumn{7}{c}{$J \to \infty$ extrapolations}    \\
\multicolumn{3}{l}{1-term eq.~(\ref{extrap1})}  & $-0.18800504$ & 0.00050504 &  8.04024  &  12.77272  \\
\multicolumn{3}{l}{2-term eq.~(\ref{extrap1})}  & $-0.18811689$ & 0.00061689 &  8.14930  &  12.95211  \\
\multicolumn{3}{l}{3-term eq.~(\ref{extrap1})}  & $-0.18817637$ & 0.00067637 &  8.23457  &  13.09609  \\  
\multicolumn{3}{l}{4-term eq.~(\ref{extrap1})}  & $-0.18821031$ & 0.00071031 &  8.30116  &  13.21039  \\ 
\end{tabular}
\end{ruledtabular}
\end{table*} 

The Hamiltonian was diagonalized in a basis constructed from a 
large number of single particle orbitals, including orbitals up 
to $\ell = 10$.  There were $20$ radial basis functions for each 
$\ell$.  Note, the symmetry of the state prevented the electrons
or positrons from occupying $\ell = 0$ orbitals.  The largest 
calculation was performed with $J = 10$ and
$L_{\rm int} = 3$ and gave a CI basis dimension of 369200.  
The parameter $L_{\rm int}$ does not have to be particularly large
since it is mainly concerned with electron-electron correlations 
\cite{bromley02b}.  The resulting Hamiltonian matrix was diagonalized 
with the Davidson algorithm \cite{stathopolous94a}, and a total 
of 300 iterations were required for the largest calculation.  The
present calculation is very slightly different from that reported
in \cite{mitroy07b}.  One of the $\ell = 10$ Laguerre functions
in \cite{mitroy07b} was input with the incorrect $n$.  The inclusion 
of the correct Laguerre function resulted in the final binding energy 
reported in \cite{mitroy07b} changing by about 1$\%$.

\begin{figure}[tbh]
\centering{
\includegraphics[width=8.8cm,angle=0]{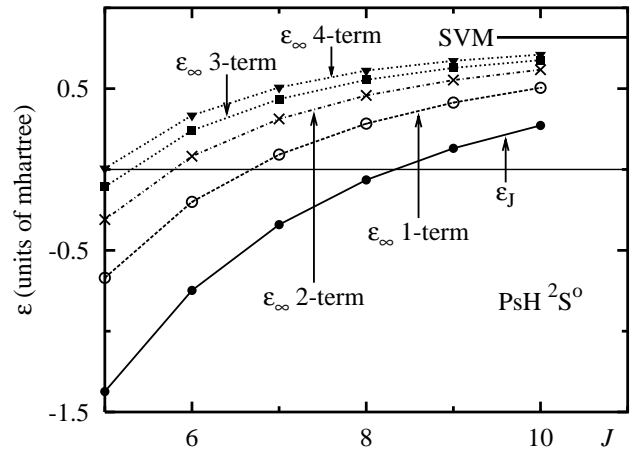}
}
\caption[]{ \label{fig:PsHE}
The binding energy, $\varepsilon = -(\langle E \rangle+0.1875)$,  
of the $^{2,4}$S$^{\rm o}$ state of PsH as a function of $J$.  
The directly calculated energy  
is shown as the solid line while the $J \to \infty$ limits using 
eq.~(\ref{extrap1}) with 1, 2, 3 or 4 terms are shown as the dashed 
lines.  The binding energy of the SVM wave function is also shown.
The H($2p$) + Ps($2p$) dissociation threshold is shown 
as the horizontal solid line.  
.
}
\end{figure}

The energy of the PsH $^{2,4}$S$^o$ state as a function of $J$ is
given in Table \ref{tab:PsH}. The calculations only give an energy 
lower than the H($2p$) + Ps($2p$) threshold of $-0.1875$ hartree 
for $J \ge 9$.  Figure \ref{fig:PsHE} shows the estimates of 
$\langle E \rangle_{\infty}$ as a function of $J$.  A quick
visual examination suggests that the extrapolations are converging 
to a common energy which attests to the reliability of the 
extrapolations in $J$.  The impact of the extrapolations is 
significant since they more than double the binding energy.  
The best CI estimate of the binding energy is the four-term 
extrapolation at $J = 10$ listed in Table \ref{tab:PsH}, namely 
$7.10 \times 10^{-4}$ hartree.  The main area 
where improvement could be made is in the dimension of the radial 
basis.  A precursor to the present CI calculation with 15 LTOs 
gave an extrapolated binding energy of $6.06 \times 10^{-4}$ hartree.   

The extrapolations of the other expectation values in Table 
\ref{tab:PsH} were done using eq.~(\ref{extrap1}).  It should
be noted that there is no formal justification for the use
of eq.~(\ref{extrap1}) for expectation values other than
the energy, so there is an additional degree of uncertainty
for these extrapolations.  In practice, this extra uncertainty 
is not that significant since the finite dimension of the radial      
basis represents a larger source of error.

\subsubsection{The stochastic variational method} 

For the PsH calculation, the two electrons have been placed
in $l_1=l_2=1$ orbitals and coupled to an $L = 1$ state 
with a total spin of $S = 1$.  The positron is then placed
into an $l_0=1$ orbital and the whole composite is coupled 
to $L=0$. The largest calculation had a total of 400 ECGs.
A summary of the energy and other expectation values is given
in Table \ref{SVM}.  

The energy of the best SVM wave function was $-0.188317$, yielding
a binding energy of $8.17 \times 10^{-4}$ hartree.  The deviation 
of the wave function from the exact virial theorem expectation, 
$(\langle V \rangle /\langle T \rangle + 2)$,  was 
$5.1 \times 10^{-8}$.  A many body system interacting by purely
coulombic interactions is known to satisfy 
$(\langle V \rangle /\langle T \rangle = -2)$ \cite{suzuki98a}.
The SVM binding energy is just 
over 10$\%$ larger than the CI energy and should be closer to 
the variational limit.   

One interesting aspect of the $^4$S$^{\rm o}$ state is that it
is more tightly bound than its H($2p^2$) parent. Its binding energy
is more that twice as large as the H($2p^2$) binding energy of
$3.55 \times 10^{-4}$ hartree \cite{bylicki03a}.  It is also
more compact.  The mean electron distance from the nucleus
of  $\langle r_e \rangle = 8.86$ $a_0$ is smaller than
that for H($2p^2$), namely $\langle r_e \rangle = 11.66$ $a_0$.
In short, the addition of the positron has resulted in a complex
that has a larger binding energy than its 3-body parent. 

The SVM radial expectation for the positron, $\langle r_p \rangle$ was
14.24 $a_0$, somewhat larger than the extrapolated CI value of 
13.67 $a_0$.  In the CI calculation the positron is localized closer 
to the nucleus
even though the CI wave function is less tightly bound.  This
is a purely computation limitation, due to the nature of LTO basis 
which is relatively compact.
Improving the radial expectation for the CI wave function would
require an increase in the number of radial functions 
per $\ell$.    
 
\begin{figure}[tbh]
\centering{
\includegraphics[width=8.8cm,angle=0]{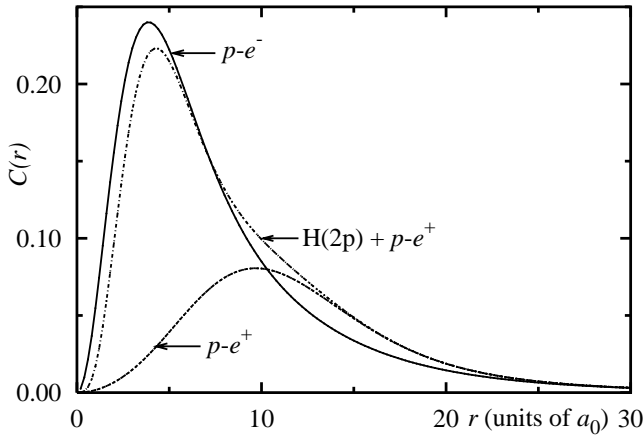}
}
\caption[]{ \label{fig:PsHCR1}
The correlation functions for the $(p,e^-)$ and $(p,e^+)$ particles
of PsH.  Also shown is a correlation function obtained by adding the
$(p,e^+)$ correlation function to the $(p,e^-)$ correlation
function of the H$(2p)$ state.  
}
\end{figure}

\begin{figure}[tbh]
\centering{
\includegraphics[width=8.8cm,angle=0]{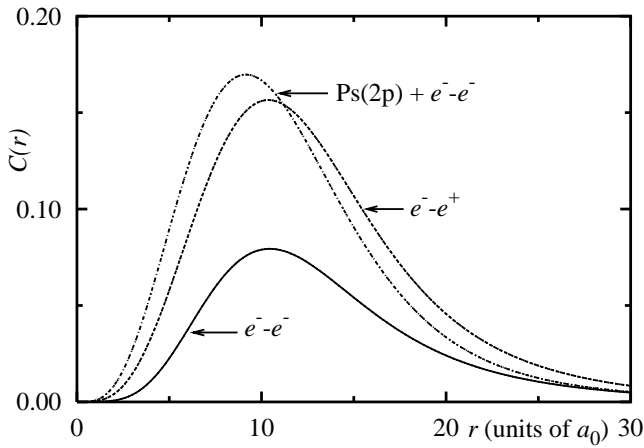}
}
\caption[]{ \label{fig:PsHCR2}
The correlation functions for the $(e^-,e^-)$ and $(e^-,e^+)$ particles 
of PsH.  Also shown is a correlation function obtained by adding the 
$(e^-,e^-)$ correlation function to the $(e^-,e^+)$ correlation function 
of the Ps$(2p)$ state.  
}
\end{figure}

The inter-particle correlation function, $C(r)$, is defined as the 
probability of finding any pair of particles a certain distance apart.  
The correlation functions shown in Figures \ref{fig:PsHCR1} and Figures 
\ref{fig:PsHCR2} are consistent with a structure consisting of a 
Ps($2p$) complex weakly bound to the H($2p$) state.  Consider an 
idealized structure consisting of a product wave function of the 
form $\Psi = \Phi(\text{Ps}[2p]) \Phi(\text{H}[2p]) \Phi_{\text{Ps}}(R)$ 
where  $\Phi_{\text{Ps}}(R)$ is the wave function describing the motion
of the Ps($2p$) center of mass. The $(p,e^-)$ and ($p,e^+)$ 
correlation functions arising from the Ps($2p$) cluster should be
the same.  Therefore, adding the ($p,e^+)$ correlation function
to the ($p,e^-$) $C(r)$ of H($2p$) state should give a correlation 
function that is the same as the actual ($p,e^-)$ correlation 
Figure \ref{fig:PsHCR1} shows a strong degree of resemblance between 
the actual ($p,e^-)$ correlation function and that obtained from a 
$\Phi(\text{Ps}[2p]) \Phi(\text{H}[2p]) \Phi_{\text{Ps}}(R)$.
Similarly, adding the ($e^-,e^-)$ correlation function to the 
($e^+,e^-$) $C(r)$ of Ps($2p$) state should give a correlation 
function that is the same as the actual ($e^-,e^-)$ correlation 
function.  Once again, the two curves shown in Figure \ref{fig:PsHCR2} 
a degree of similarity.

The energies of the finite mass variants of PsH have also 
been determined.  The energies of Ps$^1$H, Ps$^2$H, and 
Ps$^3$H are $-0.1882398$, $-0.1882784$ and $-0.1882913$ hartree
respectively.  The binding energies are 
$8.078 \times 10^{-4}$, $8.124 \times 10^{-4}$, 
$8.140 \times 10^{-4}$, hartree respectively.  

\subsection{LiPs} 

The $^2$S$^{\rm o}$ state of LiPs is a very unusual state in that it 
is a Borromean state \cite{zhukov93a,blume02a,richard03a,richard06a}. 
This is because all the possible 3-body parent states, namely the 
$^3$P$^{\rm e}$ states of Li$^-$, $e^+$Li or Ps$^-$, are themselves
unstable.  

The Ps$^-$ ion has been thoroughly investigated and does not possess 
a stable $^3$P$^{\rm e}$ state \cite{mills81b,bhatia83a}.  

Similarly, the Li$^-$ ion is believed not to have a stable
$^3$P$^{\rm e}$ state \cite{norcross74a,dulieu89b}.  We have 
also performed some very large CI calculations upon the 
Li$^-$ ion and these calculations gave no indication of
a bound state in the $^3$P$^{\rm e}$ symmetry.    

Finally, the $e^+$Li system is also not stable in the 
$^3$P$^{\rm e}$ channel.  Once again a very large CI 
calculation has been performed and once again there was 
no indication of a bound state.  Further, some calculations 
of $e^+$-Li scattering in the $^3$P$^{\rm e}$ channel also
gave no sign of a bound state.  The polarizabilities given
in Table \ref{tab0} also indicate that it should be easier
to bind a positron to the H($2p$) state than the Li($2p$).     
The SVM was also used to check whether the $e^+$H state 
is stable in the $^3$P$^{\rm e}$ channel, and once again
there was no indication of a bound state.   
   
The calculations upon LiPs were very similar in scope and scale
to those carried out upon PsH although the calculations were 
taken to $J=11$ in order to have an explicit calculation that
gave binding.  The sequence of CI energies and 
other expectation values as a function of $J$ are given in Table 
\ref{tab:LiPs}.  The binding energy $\varepsilon_J$ is defined 
as $\varepsilon_J = -(\langle E \rangle + 0.19273850)$.   

\begin{figure}[bht]
\centering{
\includegraphics[width=8.8cm,angle=0]{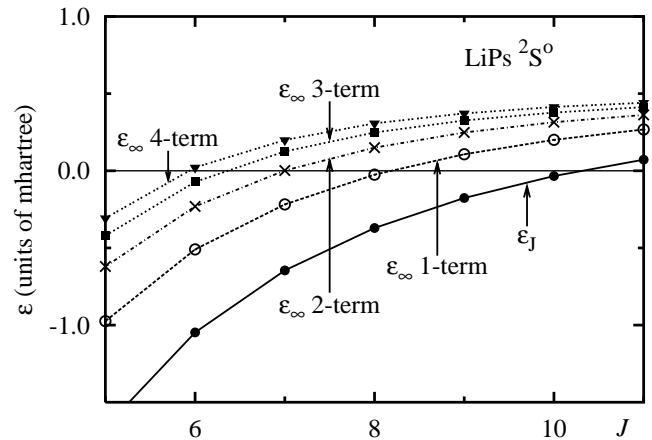}
}
\caption[]{ \label{fig:LiPsE}
The binding energy of the $^{2,4}$S$^{\rm o}$ 
state of LiPs as a function of $J$.  The directly calculated binding 
energy is shown as the solid line while the $J \to \infty$ limits using 
eq.~(\ref{extrap1}) are shown as the dashed 
lines.  The Li($2p$) + Ps($2p$) dissociation threshold is shown 
as the horizontal solid line.
lines.   
}
\end{figure}

Figure \ref{fig:LiPsE} depicts the binding energy and extrapolations 
as a function of $J$.  Only for the $J=11$ basis has 
$\langle \varepsilon \rangle_J$ crossed the threshold for binding.

\begin{table*}[th]
\caption[]{  \label{tab:LiPs}
Results of CI calculations for the $^1$S$^{\rm o}$ state of
LiPs for a series of $J$, with fixed $L_{\rm int}=3$.
The 3-body energy of the system, relative to
the energy of the Li$^{+}$ core, is denoted by $E$ (in hartree).
The threshold for binding is -0.19273850 hartree, and
$\varepsilon$ gives the binding energy (in hartree) against
dissociation into Ps($2p$) + Li($2p$).  The core annihilation rate
in units of  s$^{-1}$ is given in the $\Gamma_c$ column. 
The numbers in square brackets indicate powers of 10.
Other aspects of the table design are identical to those of 
\ref{tab:PsH}.
}
\vspace{0.2cm}
\begin{ruledtabular}
\begin{tabular}{lccccccc}
$J$&  $n$ & $N_{CI}$ & $E$ & $\varepsilon$ & $\langle r_e \rangle$ &  $\langle r_p \rangle$
                                                             & $\Gamma_c$    \\ \hline
 1  & 20  &   4200  & $-0.17291946$  & $-0.01981904$  &  6.86606 &  12.84274 & 4.5441[4]  \\ 
 2  & 40  &  16400  & $-0.18449498$  & $-0.00824351$  &  6.75226 &  11.69309 & 6.5727[4]  \\ 
 3  & 60  &  45000  & $-0.18828973$  & $-0.00444877$  &  6.89326 &  11.50143 & 6.5611[4]  \\
 4  & 80  &  85000  & $-0.19007657$  & $-0.00266192$  &  7.04675 &  11.56062 & 6.1767[4]  \\
 5  & 100 &  129200 & $-0.19107655$  & $-0.00166195$  & 7.18989 & 11.69580 & 5.7955[4] \\
 6  & 120 &  177200 & $-0.19169133$  & $-0.00104717$  & 7.32070 & 11.85451 & 5.4761[4] \\
 7  & 140 &  225200 & $-0.19209300$  & $-0.00064550$  & 7.43986 & 12.01755 & 5.2173[4] \\
 8  & 160 &  273200 & $-0.19236789$  & $-0.00037061$  & 7.54860 & 12.17700 & 5.0080[4] \\
 9  & 180 &  321200 & $-0.19256282$  & $-0.00017568$  & 7.64806 & 12.32954 & 4.8376[4] \\
 10 & 200 &  369200 & $-0.19270505$  & $-0.00003345$  & 7.73914 & 12.47367 & 4.6972[4] \\ 
 11 & 220 &  417200 & $-0.19281127$  & $0.00007278$  & 7.82273 & 12.60906 & 4.5806[4] \\ \hline 
\multicolumn{8}{c}{$J \to \infty$ extrapolations}  \\  
\multicolumn{3}{l}{1-term eq.~(\ref{extrap1})} & $-0.19300706$ & 0.000268567 &  7.97681  &  12.85861  &  4.3656[4] \\ 
\multicolumn{3}{l}{2-term eq.~(\ref{extrap1})} & $-0.19310120$ & 0.000362688 &  8.09858  &  13.06361  &  4.2226[4] \\ 
\multicolumn{3}{l}{3-term eq.~(\ref{extrap1})} & $-0.19315143$ & 0.000412934 &  8.19533  &  13.23002  &  4.1331[4] \\ 
\multicolumn{3}{l}{4-term eq.~(\ref{extrap1})} & $-0.19318032$ & 0.000441825 &  8.27254  &  13.36466  &  4.0698[4] \\ 
\end{tabular}
\end{ruledtabular}
\end{table*}

The most reliable estimates of the energy is that given after 
the 4-term extrapolation is used to determine the $J \to \infty$ 
limit of the binding energy.  The different curves in Figure 
\ref{fig:LiPsE} tend to be closer together as the number
of terms in the extrapolation increase.  The binding energy of 
$4.42 \times 10^{-4}$ hartree is just over half that of
the PsH state.

The positron can annihilate with the core electrons 
via the $2\gamma$ process since the symmetry considerations are
irrelevant here.  However, the annihilation rate of 
$\Gamma_{\rm core} \approx 4 \times 10^4$ s$^{-1}$ is small because
the positron cannot occupy a $\ell = 0$ orbital.      
  
The mean positron-nucleus distance of $\langle r_{p} \rangle = 13.5$
$a_0$ for the CI wave function was almost the same as the CI wave function 
estimate for PsH despite the smaller binding energy.  Part of the
reason for this lies in the LTO basis sets which were almost identical
for the two atoms.  The finite range of the LTO basis could be acting
to artificially confine the positron.  However, it must be remembered
that the asymptotic Ps($2p$) cluster will also be confined by the
$L = 1$ centrifugal barrier.
  
\subsubsection{The stochastic variational method} 

For the SVM, in the LiPs case, the first two electrons are placed in the 
$l_1=l_2=0$ orbits and their spins are coupled to zero, the next two
electrons are in the $l_3=l_4=1$ orbits (with the total angular momentum
coupled to 1) and their spins are coupled to 1.  Finally, the positron 
is placed in an $l_0=1$ orbit and the total orbital angular momentum 
is coupled to 1.  

The threshold for binding is the Li($2p$) ($E = -7.4101565$ hartree \cite{yan98b}) 
plus the Ps($2p$) energy.   So the energy threshold for an absolute variational
proof of binding is at $-7.4726565$ hartree.  The energy and 
expectation values of the SVM LiPs wave functions are listed in Table \ref{SVM}.  
The best variational energy was $-7.472871$ hartree, equivalent to a binding
energy of $2.15 \times 10^{-4}$ hartree.  The energy optimization was not fully 
completed and the binding energy of the CI calculation is probably more 
reliable. The primary purpose of the SVM calculation was to give an absolute 
proof that the unnatural parity state of LiPs was electronically stable.   

\subsection{NaPs} 

\begin{table*}[th]
\caption[]{  \label{tab:NaPs}
The energy of the $^{2,4}$S$^{\rm o}$ state of NaPs as a function 
of $J$.  Energies are given relative to that of the Na$^+$ core
while the threshold for binding is $-0.17406287$ hartree.  The 
column $n_-$ gives the total number of occupied electron orbitals, 
while $n_+$ gives the number of positron orbitals.  
Other aspects of the table design are identical to those of 
\ref{tab:LiPs}.
}
\begin{ruledtabular}
\begin{tabular}{lcccccccc} 
$J$&  $n_{-}$ & $n_{+}$ & $N_{CI}$ & $E$ & $\varepsilon$ & $\langle r_e \rangle$ &  $\langle r_p \rangle$
                                                             & $\Gamma_c$    \\ \hline
 1  &  21   & 20  &   4620  & $-0.15378569$ & $-0.02027718$ &  7.74287 &  13.88608 &  1.4441[5] \\  
 2  &  41   & 40  &   17220 & $-0.16614255$ & $-0.00792032$ &  7.58513 &  12.48950 &  2.2339[5] \\
 3  &  61   & 60  &   46220 & $-0.17033251$ & $-0.00373035$ &  7.71039 &  12.16854 &  2.2948[5] \\
 4  &  81   & 80  &   86620 & $-0.17231600$ & $-0.00174687$ &  7.85240 &  12.16161 &  2.1849[5] \\
 5  & 101 &  100  &  131220 & $-0.17341763$ & $-0.00064523$ &  7.98365 &  12.25289 &  2.0625[5] \\
 6  & 121 &  120  &  179620 & $-0.17408661$ &  0.00002375 &  8.10085 &  12.37548  &  1.9577[5] \\
 7  & 141 &  140  &  228020 & $-0.17451658$ &  0.00045371 &  8.20440 &  12.50450  &  1.8733[5] \\
 8  & 161 &  160  &  276420 & $-0.17480552$ &  0.00074266 &  8.29583 &  12.62983  &  1.8060[5] \\
 9  & 181 &  180  &  324820 & $-0.17500636$ &  0.00094349 &  8.37645 &  12.74713  &  1.7523[5] \\ 
 10 & 201 &  200  &  373220 & $-0.17514972$ &  0.00108685 &  8.44734 &  12.85449  &  1.7093[5] \\ \hline  
           \multicolumn{9}{c}{$J \to \infty$ extrapolations}      \\
\multicolumn{4}{l}{1-term eq.~(\ref{extrap1})}   & $-0.17538587$ & 0.00132300 & 8.56412 &  13.03133 &  1.63839[5] \\
\multicolumn{4}{l}{2-term eq.~(\ref{extrap1})}   & $-0.17549381$ & 0.00143094 & 8.65082 &  13.17118 &  1.59424[5] \\ 
\multicolumn{4}{l}{3-term eq.~(\ref{extrap1})}   & $-0.17554787$ & 0.00148500 & 8.71500 &  13.27807 &  1.56638[5] \\
\multicolumn{4}{l}{4-term eq.~(\ref{extrap1})}   & $-0.17557663$ & 0.00151376 & 8.76225 &  13.35833 &  1.54841[5] \\ 
\end{tabular}
\end{ruledtabular}
\end{table*}

The calculations upon NaPs were very similar in scope and scale
to those carried out upon LiPs.  About the only difference was
that an extra $\ell=1$ orbital was added to the electron basis.

The energies of the Na($3s$) and Na($3p$) states in the 
model potential were $-0.18885491$ and $-0.11156287$ hartree.  The
experimental binding energies are $-0.188858$ and $-0.111547$   
hartree respectively \cite{nistasd3}.  Electronic stability 
requires a total 3-body energy of $-0.17406287$ hartree and  
the binding energy $\varepsilon_J$
is defined as $\varepsilon_J = -(\langle E \rangle + 0.17406287)$.   
The energy of the $^3$P$^{\rm e}$ excited state of Na$^-$
is $-0.11342529$ hartree, i.e the Na($3p$) has an electron affinity
of 0.002262 hartree with respect to attaching an electron to
the $^3$P$^{\rm e}$ state.  This is reasonably close to the
original value of Norcross, 0.00228 hartree \cite{norcross74a}.

\begin{figure}[bht]
\centering{
\includegraphics[width=8.8cm,angle=0]{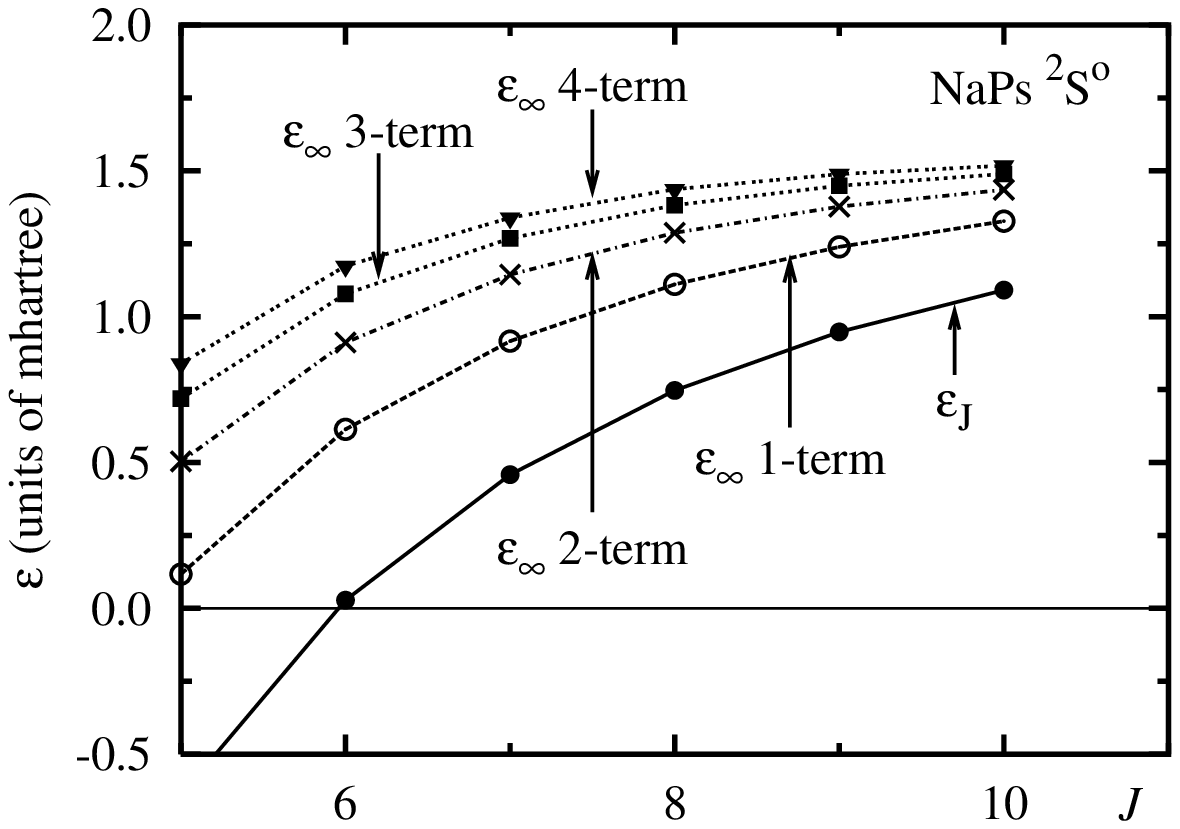}
}
\caption[]{ \label{fig:NaPsE}
The binding energy of the $^{2,4}$S$^{\rm o}$ 
state of NaPs as a function of $J$.  The directly calculated binding 
energy is shown as the solid line while the $J \to \infty$ limits using 
eq.~(\ref{extrap1}) with 1, 2 or 3 terms are shown as the dashed 
lines.  The Na($3p$) + Ps($2p$) dissociation threshold is shown 
as the horizontal solid line.
}
\end{figure}

Table \ref{tab:NaPs} gives the energies and radial expectation
values  as a function of $J$ while figure \ref{fig:NaPsE} shows 
the variation of $\varepsilon_{\infty}$ as a function of $J$.  
Once again the three and four term extrapolations seem to be 
converging to a common energy.   In this case the $J \to \infty$ 
correction increases the binding energy by about 40 $\%$ from 
$10.87 \times 10^{-4}$ hartree to $15.14 \times 10^{-4}$ hartree.
The binding energy of the NaPs unnatural parity state is about 
twice as large as that of PsH.  

The annihilation rate with the core electrons was 
$\Gamma_{\rm core} \approx 1.5 \times 10^5$ s$^{-1}$ is small.
Although this is 3 times larger than $\Gamma_{\rm core}$ for LiPs,
in absolute terms the annihilation rate is still small.
  
\subsection{KPs} 

\begin{table*}[th]
\caption[]{  \label{tab:KPs}
The energy of the $^{2,4}$S$^{\rm o}$ states of KPs as a function 
of $J$.  The threshold for binding is $-0.16268265$ hartree and  
and the energies are given relative to that of the K$^+$ core.
Other aspects of the table design are identical to those of 
Table \ref{tab:NaPs}.
}
\begin{ruledtabular}
\begin{tabular}{lcccccccc} 
$J$ &  $n_{-}$ & $n_{+}$ & $N_{CI}$ & $E$ & $\varepsilon$ & $\langle r_e \rangle$ &  $\langle r_p \rangle$
                                                             & $\Gamma_c$    \\ \hline
  1 &  21 &  20 &   4620 & -0.14168301 & -0.02099965 & 8.49503 & 14.71924 & 3.4488[5] \\
  2 &  41 &  40 &  17220 & -0.15457885 & -0.00810380 & 8.29108 & 13.13410 & 5.4508[5] \\
  3 &  61 &  60 &  46220 & -0.15914469 & -0.00353796 & 8.39896 & 12.70578 & 5.7120[5] \\
  4 &  81 &  80 &  86620 & -0.16132591 & -0.00135675 & 8.53370 & 12.64979 & 5.4824[5] \\
  5 & 101 & 100 & 131220 & -0.16253700 & -0.00014566 & 8.66023 & 12.71399 & 5.1917[5] \\
  6 & 121 & 120 & 179620 & -0.16326995 &  0.00058729 & 8.77291 & 12.81839 & 4.9358[5] \\
  7 & 141 & 140 & 228020 & -0.16373804 &  0.00105538 & 8.87154 & 12.93315 & 4.7280[5] \\
  8 & 161 & 160 & 276420 & -0.16405038 &  0.00136772 & 8.95749 & 13.04566 & 4.5629[5] \\
  9 & 181 & 180 & 324820 & -0.16426580 &  0.00158314 & 9.03227 & 13.15085 & 4.4319[5] \\
 10 & 201 & 200 & 373220 & -0.16441830 &  0.00173564 & 9.09676 & 13.24589 & 4.3282[5] \\ \hline 
           \multicolumn{9}{c}{$J \to \infty$ extrapolations}      \\
\multicolumn{4}{l}{1-term eq.~(\ref{extrap1})}  & $-0.16466950$ &  0.00198684 &  9.20300  &  13.40245 & 4.1575[5] \\  
\multicolumn{4}{l}{2-term eq.~(\ref{extrap1})}  & $-0.16478114$ &  0.00209848 &  9.27933  &  13.52387 & 4.0541[5] \\ 
\multicolumn{4}{l}{3-term eq.~(\ref{extrap1})}  & $-0.16483516$ &  0.00215250 &  9.33339  &  13.61318 & 3.9924[5] \\ 
\multicolumn{4}{l}{4-term eq.~(\ref{extrap1})}  & $-0.16486273$ &  0.00218008 &  9.37076  &  13.67636 & 3.9562[5] \\  
\end{tabular}
\end{ruledtabular}
\end{table*} 

The calculations upon KPs were very similar in scope and scale
to those carried out upon LiPs.  About the only difference was
that an extra $\ell=1$ orbital was added to the electron basis.

The energies of the K($4s$) and K($4p$) states in the 
model potential were $-0.159520$ and $-0.10018265$ hartree.  The
experimental binding energies are $-0.159516$ and $-0.100176$   
hartree respectively \cite{nistasd3}.  Electronic stability 
requires a total 3-body energy of $-0.16268265$ hartree and  
the binding energy $\varepsilon_J$
is defined as $\varepsilon_J = -(\langle E \rangle + 0.16268265)$.   
The energy of the $^3$P$^{\rm e}$ excited state of K$^-$
is $-0.104498$ hartree, i.e the K($4p$) has an electron affinity
of 0.004322 hartree with respect to attaching an electron to
the $^3$P$^{\rm e}$ state.  This is close to the original 
value of Norcross, 0.00437 hartree \cite{norcross74a}.

\begin{figure}[bht]
\centering{
\includegraphics[width=8.8cm,angle=0]{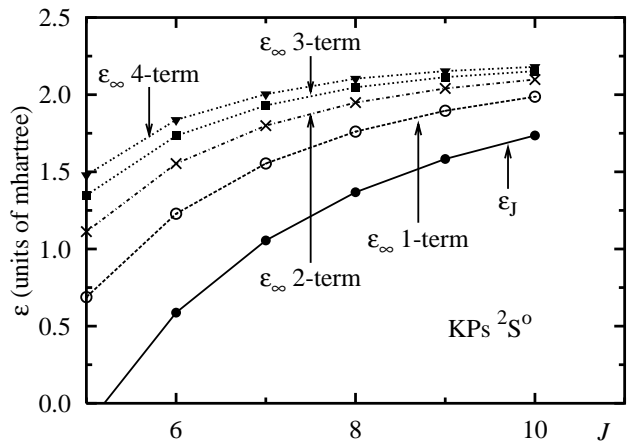}
}
\caption[]{ \label{fig:KPsE}
The binding energy of the $^{2,4}$S$^{\rm o}$ 
state of KPs as a function of $J$.  The directly calculated binding 
energy is shown as the solid line while the $J \to \infty$ limits using 
eq.~(\ref{extrap1}) are shown as the dashed 
lines.  The K($4p$) + Ps($2p$) dissociation threshold is shown 
as the horizontal solid line.
lines.   
}
\end{figure}

Table \ref{tab:KPs} gives the energies and radial expectation
values as a function of $J$ while figure \ref{fig:KPsE} shows 
the variation of $\varepsilon_{\infty}$ as a function of $J$.  
The three and four term extrapolations seem to be converging 
to a common energy.   In this case the $J \to \infty$ 
corrections increase the binding energy by about 20$\%$ from 
$17.36 \times 10^{-4}$ hartree to $21.80 \times 10^{-4}$ hartree.
The KPs system has the largest binding energy of all the systems
considered in this paper.  

\section{Summary}

A number of PsX systems (X = H, Li, Na and K) are seen to have
electronically stable $^{2,4}$S$^{\text o}$ complexes that are 
stable against auto-ionization, and in addition these states
only decay slowly by positron annihilation.  All the particles
in these effectively four-body complexes are in a relative $P$-state 
with respect to each other.  The most unusual of the systems is 
LiPs since the $^{2,4}$S$^{\text o}$ states are of Borromean 
type.   The sequence of calculations suggest that there would
also exist unnatural parity $^{2,4}$S$^{\rm o}$ complexes of
RbPs and CsPs; and most likely they would have binding energies
larger than KPs.    

Due to their low binding energies, these systems can be expected to 
have a structure composed of an Ps($2p$) cluster loosely bound to an 
atomic X($np$) excited state.  This has been confirmed by the 
correlation functions for PsH which were computed using the SVM.  
 
Although these complexes are electronically stable and decay very
slowly by electron-positron annihilation there are other decay processes 
that act to shorten the lifetime.  These complexes can emit a photon, 
decaying to a state of  $^{2,4}$P$^{\text e}$ symmetry.  For example, 
a Ps($np$) fragment in the complex can emit a photon decaying to a 
Ps($1s$) type fragment.  The Ps($1s$) fragment could then annihilate 
by the $2\gamma$ or $3\gamma$ process.  In addition, the resulting 
$^{2,4}$P$^{\text e}$ state could also decay by auto-ionization.  
The lifetime of these states can be expected to be comparable to 
the lifetime of the fragments against single photon decay, e.g. 
X($2p$) $\to$ X($1s$).  So the overall lifetimes of the states can 
be expected to be of order $10^{-8}$ - $10^{-9}$ seconds.    

It is unlikely that any of these complexes will be identified in
the laboratory in the near future.  The formation of positronic 
compounds is known to be notoriously difficult \cite{charlton01b}.  
That these states are unnatural parity states compounds the 
difficulty since such states are not readily formed in normal 
collision systems.  For example, the $^3$P$^{\rm e}$ ion states 
\cite{holoien61a,drake70a,norcross74a,bylicki03a} that could serve
as suitable parents have never been identified in the laboratory.

Besides the PsH and APs systems, there are other related physical 
systems that could have unnatural parity bound states.  For example, 
there is the possible existence of a new bi-exciton excited state 
\cite{usukura99a}.  While the Ps$^-$ ion might not have a stable 
$^3$P$^{\text e}$ state, it is known that the ($M^+$, $e^-$, $e^-$) 
ion is stable for $M^+/m_e < 0.4047$ and $M^+/m_e > 16.8$ 
\cite{mills81b,bhatia83a}.  It could be expected that a bi-exciton 
state, ($e^-$,$e^-$,$h$, $h$), with $^{1,3,5}$S$^{\text o}$ symmetry 
would be electronically stable when the mass ratios make the 
$^3$P$^{\text e}$ state of the charged exciton ($e^-$,$e^-$,$h$) 
stable.  The system might also exhibit Borromean binding, there
might be a bound bi-exciton state even though neither of the
$^3$P$^{\text e}$ ($e^-$,$e^-$,$h$) or ($e^-$,$h$,$h$) states
was stable.  
    
\begin{acknowledgments}

These calculations were performed on Linux clusters hosted at the 
South Australian Partnership for Advanced Computing (SAPAC) and 
SDSU Computational Sciences Research Center, with system 
administration given by Grant Ward, Patrick Fitzhenry and Dr James Otto.  
This work in part was supported by NSF grant ECS 0622146.  

\end{acknowledgments}


\end{document}